# SECURING GENERATIVE AI AGENTIC WORKFLOWS: RISKS, MITIGATION, AND A PROPOSED FIREWALL ARCHITECTURE

**SUNIL KUMAR JANG BAHADUR[1], GOPALA DHAR[2]**

[1]AI & GenAI Specialist, Cloud GTM, Google, Mumbai, India
[2]AI Engineer, AI Services, Google Cloud Consulting (GCC), Google. Mumbai, India
E-MAIL: sjangabahadur@google.com, gopalad@google.com

**Abstract:**

Generative Artificial Intelligence (GenAI) presents significant advancements but also introduces novel security challenges, particularly within agentic workflows where AI agents operate autonomously. These risks escalate in multi-agent systems due to increased interaction complexity. This paper outlines critical security vulnerabilities inherent in GenAI agentic workflows, including data privacy breaches, model manipulation, and issues related to agent autonomy and system integration. It discusses key mitigation strategies such as data encryption, access control, prompt engineering, model monitoring, agent sandboxing, and security audits. Furthermore, it details a proposed "GenAI Security Firewall" architecture designed to provide comprehensive, adaptable, and efficient protection for these systems by integrating various security services and leveraging GenAI itself for enhanced defense. Addressing these security concerns is paramount for the responsible and safe deployment of this transformative technology.

**Keywords:**
GenAI, Security, LLM, Agents, Firewall

## 1. Introduction

Generative AI technologies are rapidly evolving, enabling sophisticated applications through agentic workflows where AI agents execute tasks autonomously. While powerful, these workflows create a new attack surface for malicious actors. The complexity and potential for unforeseen vulnerabilities are amplified in multi-agentic systems, where multiple agents interact, increasing the risk of unintended consequences. Ensuring the security and integrity of these systems necessitates a multi-layered approach, encompassing robust data security, rigorous model validation, ethical development guidelines, and continuous monitoring. As GenAI adoption grows, proactively addressing the associated security challenges is crucial for harnessing its benefits safely and responsibly. This paper focuses specifically on identifying the security risks pertinent to GenAI-based agentic workflows and proposes a comprehensive solution architecture to mitigate these threats.

## 2. Problem Statement

Implementing GenAI-based agentic workflows introduces several critical security concerns that developers and users must address. These can be categorized as follows:

### 2.1. Data Privacy and Confidentiality

Agentic workflows often process sensitive data, making protection essential throughout the lifecycle. Key risks include:

- Data Leakage: Inadvertent exposure of confidential information by inadequately secured agents.
- Data Misuse: Unauthorized access and manipulation of data by malicious actors or compromised agents.
- Compliance Violations: Non-adherence to data privacy regulations (e.g., GDPR, CCPA), leading to legal and financial penalties.

### 2.2. Model Vulnerabilities

The core GenAI models are susceptible to attacks that can compromise the entire workflow. Examples include:

- Prompt Injection: Malicious inputs designed to manipulate the model's output or behavior.
- Model Evasion: Inputs crafted to bypass security filters and controls implemented within the model or workflow.
- Model Poisoning: Introducing corrupted data into the training set to degrade model performance or embed security flaws.

- Model Theft: Stealing the model, its architecture, or training data, enabling misuse for generating fake content or launching cyberattacks.

### 2.3. Agent Autonomy and Control

The autonomous nature of agents presents unique challenges.
- Rogue Agents: Agents deviating from their intended programming to perform unauthorized or harmful actions.
- Lack of Transparency: Difficulty in tracing and understanding the decision-making process within complex, multi-agent systems.
- Escalation of Privileges: Agents potentially gaining unintended access levels to sensitive systems or data.

### 2.4. Network and System Security

Interactions with external systems and network communications introduce further risks.
- API Security: Exploitable vulnerabilities in APIs used by agents for communication or task execution.
- Network Communication: Interception of unencrypted data transmitted between agents or systems.
- System Integration: Introduction of new security flaws when integrating agentic workflows with existing IT infrastructure.

## 3. Problem Statement

Addressing the identified security issues requires implementing a combination of technical and procedural safeguards. Key strategies include:
- Data Encryption: Protecting sensitive data both at rest and in transit using strong encryption methods.
- Access Control: Implementing strict authentication mechanisms and role-based access controls to limit data and system access.
- Prompt Engineering: Carefully designing, validating, and sanitizing prompts to prevent injection attacks and manipulation.
- Model Monitoring: Continuously observing model behavior for anomalies, potential attacks, or deviations from expected performance.
- Agent Sandboxing: Isolating agents within secure, controlled environments to limit the potential impact of a compromised or rogue agent.
- Security Audits: Conducting regular vulnerability assessments and security audits of the models, workflows, and supporting infrastructure.

The table below maps specific issues to primary mitigation techniques:

**TABLE 1.** Table type styles

| Security Issue | Mitigation Strategy |
| --- | --- |
| Data Leakage | Data Encryption, Access Control |
| Prompt Injection | Prompt Engineering, Model Monitoring |
| Rogue Agents | Agent Sandboxing, Security Audits |
| API Security | Secure API Design, Regular Testing |

Furthermore, GenAI itself can be leveraged to enhance these mitigation strategies. For instance, GenAI can assist in identifying data requiring encryption, developing intelligent access control systems, refining prompts against manipulation, creating sophisticated model monitoring systems, managing sandboxing environments, and automating parts of the security audit process. Incorporating GenAI into security practices can significantly bolster defenses for agentic workflows.

Integrating an input/output scanner security layer for every Large Language Model (LLM) presents scalability challenges. This approach increases overall overhead, latency, and negatively impacts performance. Consequently, an independent, centralized firewall-like security layer capable of integrating with diverse multi-agent workflows is necessary.

## 4. Proposed Solution: GenAI Security Firewall

The proposed "GenAI Security Firewall" architecture offers a strong, central security solution for GenAI agentic workflows. As illustrated in the accompanying diagram, this system functions as a distinct service layer, providing protection for potentially multi-agent and self-optimizing systems.

### 4.1. Architecture Overview

The proposed architecture comprises several key components operating within an overall Agentic Workflow Context (defined by Core Persona, Use Case Details, Evaluation Samples, and Preferences):

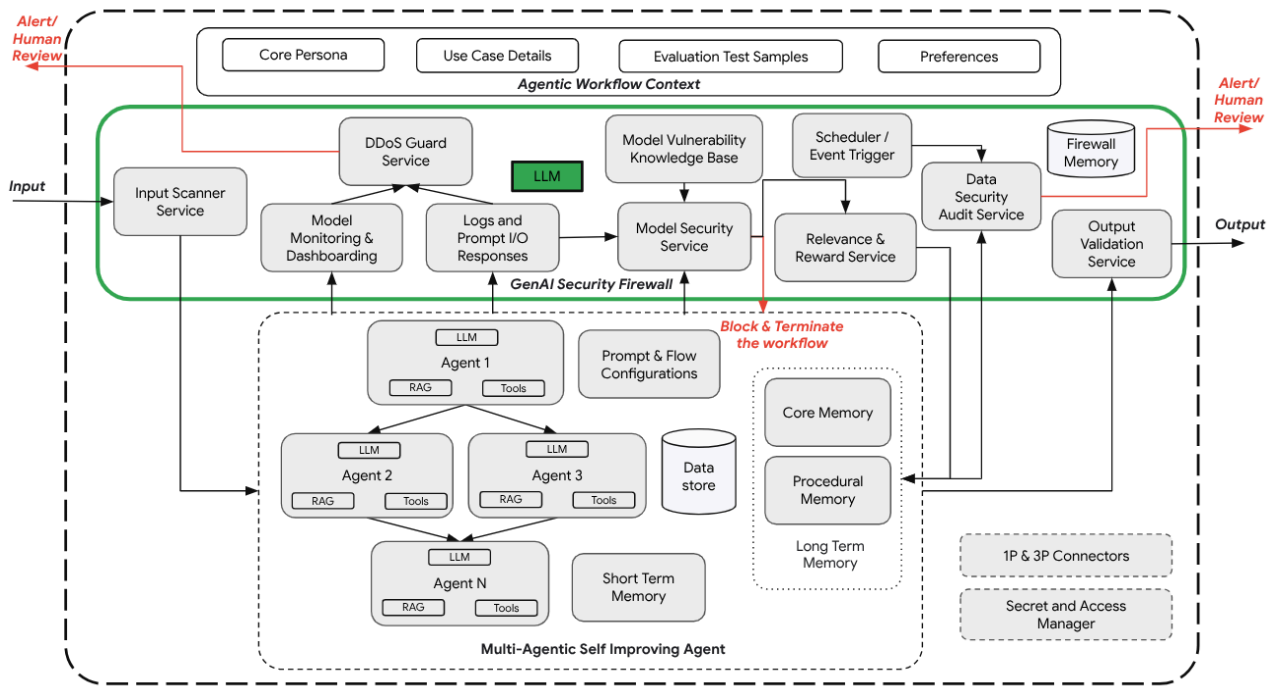

- GenAI Security Firewall:
  This core component actively monitors and protects the workflow. Its sub-components include:
    - Input Scanner Service: Scrutinizes incoming data for threats like malicious code or prompt injection.
    - DDoS Guard Service: Protects against denial-of-service attacks.
    - Model Monitoring & Dashboarding: Tracks model performance for anomalies indicating potential breaches.
    - Logs and Prompt I/O Responses: Records interactions for auditing and forensic analysis.
    - Model Vulnerability Knowledge Base: Maintains a database of known AI model vulnerabilities.
    - Model Security Service: Focuses specifically on detecting and preventing LLM-specific attacks like prompt injection.
    - Data Security Audit Service: Periodically audits stored data for integrity and malicious content.
    - Relevance & Reward Service: Evaluates response quality, potentially using ML to improve threat discernment over time by feeding insights into long-term memory.
    - Output Validation Service: Checks generated outputs for security and accuracy before release.
    - Scheduler/Event Trigger: Automates routine security tasks like audits.
    - Firewall Memory: Stores historical security event data to adapt to evolving threats.
- Multi-Agentic Self Improving Agent:
  The core workflow system being protected, potentially consisting of multiple LLM agents using Retrieval-Augmented Generation (RAG) and other tools. It includes various memory stores (Long Term, Core, Procedural, Short Term) and data stores configured by prompts and flow definitions. Long-Term Memory serves as a persistent repository for foundational knowledge and past experiences. Core Memory holds essential contextual information relevant to the current workflow. Procedural Memory stores learned sequences of actions and execution strategies. Short-Term Memory maintains immediate contextual details and temporary data necessary for ongoing processing. The management of different memory types, and the interaction with data stores, is dynamically orchestrated through prompts and meticulously defined workflows, ensuring a flexible and adaptable operational framework.

- Supporting Components:
  - Connectors (1P & 3P): Facilitate interaction with other internal and external systems.
  - Secret and Access Manager: Manages credentials and access permissions securely.

### 4.2. Workflow

The security process follows these general steps:
- Input is received and undergoes initial scanning for obvious anomalies or known malicious patterns.
- The input is then sent to the GenAI Firewall for in-depth security analysis.
- The GenAI Firewall utilizes various security services, including content filtering, threat intelligence, vulnerability detection, and behavioral analysis.
- The GenAI Firewall can interact with a multi-agent system to gain context, retrieve information, or delegate sub-tasks.
- After analysis, the GenAI Firewall categorizes the input as safe or potentially harmful.
- Safe input proceeds to output validation to ensure compliance and absence of malicious content.
- Validated output is released.
- A feedback mechanism tracks the relevance and reward of the output for continuous system improvement, refining policies and detection accuracy.
- Alerts are generated for security personnel when human oversight is needed.
- Definitive threats detected by the GenAI Firewall trigger a blocking mechanism to halt processing and prevent harm.

### 4.3. Key Features

The proposed GenAI Security Firewall offers several advantages:
- Comprehensive Security: Provides end-to-end protection by analyzing inputs, monitoring models, auditing data, and validating outputs.
- Flexible and Adaptable: The modular design, use of LLMs/RAG, and learning capabilities allow the system to adapt to new threats and contexts.
- Cost-Effective and Efficient: As an independent service layer, it can be more efficient and less latent compared to embedding security checks directly within every prompt or agent interaction.

### 4.4. Benefits of a Centralized GenAI Firewall:

- Enhanced Security Posture:
  - Reduced Policy Drift: Centralized policy enforcement can decrease policy drift by 60-80% compared to decentralized approaches, leading to fewer agents operating with outdated or conflicting security rules.
  - Reduced Attack Surface: Blocking common attack vectors centrally can prevent 80-90% of prevalent attacks like prompt injection.
  - Enhanced Threat Detection: A holistic view of the core workflow allows the firewall to identify attacks manifesting across multiple agent interactions, detecting anomalies missed by individual agents.
  - Safeguarding Data Disclosure: Controls access to GenAI applications and monitors activities to prevent unauthorized data transfers.
  - Zero Trust Implementation: Continuously authenticates and authorizes all AI interactions, minimizing unauthorized access risks.
- Improved Operational Efficiency and Cost Savings:
  - Reduced Redundancy & Cost Savings: Eliminating redundant security logic in individual wrappers can lead to 15-25% savings in maintenance costs.
  - Improved Efficiency: Centralized logging and monitoring can reduce incident response time by at least an average of 30-40%.

## 5. Conclusion

Generative AI agentic workflows hold immense potential but introduce significant security risks that cannot be ignored. The vulnerabilities span data privacy, model integrity, agent control, and system interactions. While standard mitigation strategies like encryption, access control, and monitoring are essential, the unique nature of GenAI necessitates specialized defenses.

The proposed GenAI Security Firewall architecture offers a dedicated, comprehensive approach to securing these workflows. By integrating multiple security services – from input scanning and DDoS protection to model monitoring, data auditing, and output validation – it aims to create a robust defense layer. Leveraging GenAI within the firewall itself for tasks like anomaly detection and threat analysis further enhances its capabilities. The system's adaptability, enabled by components like the vulnerability knowledge base and feedback mechanisms, is crucial for addressing the rapidly evolving threat landscape.

Implementing such robust security measures is not merely recommended but essential for the trustworthy and reliable deployment of GenAI-based agentic systems. Continued research and development in areas like model robustness, explainability for agent actions, and secure multi-agent coordination will be vital as this technology matures. By prioritizing security alongside innovation, we can ensure that GenAI agentic workflows are deployed safely and effectively.

**References**

[1] Salman Rahman, Liwei Jiang, James Shiffer, Genglin Liu, Sheriff Issaka, Md Rizwan Parvez, Hamid Palangi, Kai-Wei Chang, Yejin Choi, Saadia Gabriel, "X-Teaming: Multi-Turn Jailbreaks and Defenses with Adaptive Multi-Agents", arXiv:2504.13203, 15 Apr 2025

[2] Taicheng Guo, Xiuying Chen, Yaqi Wang, Ruidi Chang, Shichao Pei, Nitesh V. Chawla, Olaf Wiest, Xiangliang Zhang, "Large Language Model based Multi-Agents: A Survey of Progress and Challenges", arXiv:2402.01680, 19 Apr 2024

[3] Yagmur Yigit, William J Buchanan, Madjid G Tehrani, Leandros Maglaras, "Review of Generative AI Methods in Cybersecurity", arXiv:2403.08701, 19 Mar 2024

[4] Ajay Sivakumar, Shalini, Vasantha Raj, Sebastian Sylvester, "The Self-Learning Agent with a Progressive Neural Network Integrated Transformer", arXiv:2504.02489, 3 Apr 2025

[5] Wujiang Xu, Kai Mei, Hang Gao, Juntao Tan, Zujie Liang, Yongfeng Zhang, "A-MEM: Agentic Memory for LLM Agents", arXiv:2502.12110, 18 Apr 2025